\documentclass[sn-mathphys-num]{sn-jnl}

\usepackage{graphicx}%
\usepackage{multirow}%
\usepackage{amsmath,amssymb,amsfonts}%
\usepackage{slashed}
\usepackage{amsthm}%
\usepackage{mathrsfs}%
\usepackage[title]{appendix}%
\usepackage{xcolor}%
\usepackage{textcomp}%
\usepackage{manyfoot}%
\usepackage{booktabs}%
\usepackage{algorithm}%
\usepackage{algorithmicx}%
\usepackage{algpseudocode}%
\usepackage{listings}%
\usepackage{MnSymbol}
\theoremstyle{thmstyleone}%
\theoremstyle{thmstyletwo}%

\theoremstyle{thmstylethree}%

\begin{document}

\title[Article Title]{\bf{Green's functions under magnetic effects in a nontrivial topology}}

\author*[1]{\fnm{Emerson B. S.} \sur{Corrêa}}\email{ecorreae@ufpa.br}

\author[2]{\fnm{Michelli S. R.} \sur{Sarges}}\email{michellissarges@gmail.com}

\author[3]{\fnm{José A.}
\sur{Helay\"el-Neto}}\email{helayel@cbpf.br}

\affil*[1]{\orgdiv{Faculdade de Física}, \orgname{Universidade Federal do Pará - UFPA}, %

\postcode{66075-110},
\city{Belém}, \state{PA}, \country{Brazil}}


\affil[2]{\orgdiv{Independent researcher},  \city{Belém}, \state{PA}, \country{Brazil}}

\affil[3]{\orgdiv{}
\orgname{Centro Brasileiro de Pesquisas Físicas - CBPF},
\postcode{22290-180}
\city{Rio de Janeiro}, \state{RJ}, \country{Brazil}
}
\equalcont{These authors contributed equally to this work.}


\abstract{

We calculate the Bose and Dirac field propagators in a four-dimensional Euclidean space under a magnetic external field by using a hybrid version of the Ritus and Schwinger methods. We get both propagators explicitly in the coordinate and momentum domains without dimensional reduction. Through gauge transformations, we eliminate the translation non-invariant part of the propagators. Also, the Matsubara frequencies of the fields are obtained in a toroidal topology. The approach we consider in this paper has the advantage of taking into account thermal and finite-volume effects with several kinds of boundary conditions, in addition to considering all Landau levels simultaneously, in an analytical and tractable way. 

}




\maketitle


\section{Introduction}

    Magnetic fields play one of the most important physical scenarios in which quantum fields can be immersed. For example, it is believed that magnetic fields were present microseconds after the Big Bang, where a phase transition from the so-called quark-gluon plasma phase (QGP) to the hadronic phase could be occurred in the primordial Universe~\cite{AstroP,Grasso,DiagramaQCDMag,PRD50,PRD2013,DiagramaQCDMagItalianos}. There are many interesting phenomena involving magnetic backgrounds, such as the chiral magnetic effect, which one the external magnetic field creates a transport charge induced by chirality imbalance of the considered medium~\cite{MagChiral,Nature,MagChiralRev}, and the magnetic catalysis, i.e., the stimulation of the chiral broken phase, as the values of the external magnetic field become larger~\cite{Debora,Italianos1,Indianos,Kojo,PRD2024}. Oppositely, lattice quantum chromodynamics computations have shown that magnetic field intensity growth can also contribute to restoring the chiral symmetry due to the decrease of chiral condensate. This behavior (knowing as inverse magnetic catalysis) was discovered in 2012 and still draws the attention of the community nowadays~\cite{IMC-JHEP,IMC-PRD,IMC1,IMC2,IMC3,IMC4,IMC5}. Magnetic fields in the orthogonal direction to the collision plane are created in the context of non-central relativistic heavy-ion collisions. The relative motion of ions can generate large magnitudes, usually between $10^{15}-10^{16}-\mathrm{Tesla}$~\cite{IonsPesados}. For an numerical comparation, the pion mass $m_{\pi}=0.138~\mathrm{GeV}$ is relationed by a magnetic intensity of order $eB \sim m^{2}_{\pi} \sim 10^{14}\,\mathrm{Tesla}$.

Green's functions determine important properties of the corresponding quantum system analysed, such as energy levels and probability amplitudes~\cite{Farina}. In particular, considering Quantum Field Theory (QFT), Green's functions are practically indispensable in describing quantum fields that interact with an external electromagnetic field, such as in pair production rate scenarios~\cite{Schwinger,Dittrich}.  Being a suitable mathematical tool for describing the effects of interactions on the propagation of particles, there are many methods to compute Green's functions (Feynman propagators)~\cite{Schwinger,Ritus,Ritus2,Chodos,Lawrie,Dittrich,ElizaldePropRitus,Mexicanos,Igor,RBEFEmerson2015,CJPEmerson2022,CMEmerson2022}. There is additional interest in getting the Green's function when a field is heated, thanks to the possibility of using it in several kinds of description, such as in the effects coming from the thermal bath in high-energy heavy-ion collisions, in phase transition phenomena, or in the breaking or restoration of symmetry~\cite{Zinn-Justin,Zinn-Justin1,Ultracold,LivroAdolfo2009}. 

In this thermal scenario, the Matsubara formalism has been used as a powerful tool in theoretical physics since a pioneering paper written by T. Matsubara in 1955~\cite{Matsubara}. Matsubara proposed a new method to compute the thermal Green's function of a statistical system using extensively the operator technique commonly used in QFT. Then, he applied that formalism to describe an electron-phonon system, since he had obtained the thermal Green's function for bosons and fermions. The thermal distributions for bosons and fermions can be obtained by the so-called Kubo-Martin-Schwinger conditions (KMS)~\cite{Kubo,Martin-Schwinger}, which establish periodic boundary conditions on the imaginary time coordinate for bosons but antiperiodic boundary conditions for fermions at the same coordinate. As a direct consequence this periodicity (antiperiodicity) on the imaginary time coordinate $\tau$, we get discrete frequencies such that $\omega_{n_\tau} = (2 \,n_{\tau})\pi/\beta$, for bosons [or $\omega_{n_\tau} = (2 \,n_{\tau}+1)\pi/\beta$, for fermions], being $n_{\tau}$ $\in \,\mathbb{Z}$ and $\beta$ the inverse of system temperature~\cite{Dolan}. 

In 1980, the authors of the Ref.~\cite{Ford} extended the Matsubara formalism to one spatial coordinate. It was the starting point for the development of quantum field theory in a nontrivial topology and allowed us getting account for the finite volume effects on the system. However, the physical criteria prescribed by KMS conditions at the imaginary time coordinate do not have reason to be applied too at spatial coordinates. Actually, at this point, we have the freedom to choose what kind of boundary condition to use at the spatial coordinate. For instance, the authors of Ref.~\cite{PhysRep2014} used periodic boundary conditions in the bosonic systems in $(D-1)$ spatial coordinates but anti-periodic boundary conditions in their counterpart fermionic systems. Notwithstanding, in Ref.~\cite{Inagaki2022}, phase diagrams are obtained using the effective potential technique in low dimensions, considering periodic and anti-periodic boundary conditions over the same spatial dimension compacted in a fermionic system. In the literature, we find several papers with periodic, anti-periodic, and twisted boundary conditions in spatial coordinates of the system, no matter if the quantum field has integer or half spin, see Refs.~\cite{twisted,Erich,NuclPhysAEmerson2023,EPJPlus}.

When the physical system is embedded in an external magnetic background, we observe the \lq\lq mathematical phenomenon\rq\rq \,dimensional reduction ~\cite{MC1-PRL,MC1,Kharzeev,Miransky,IMC66}. In this case, the system has two dimensions hidden due to the external field applied. Thus, a typical system in interaction defined in four dimensions, with coordinates in Euclidean space given by $(\tau,x,y,z)$, becomes after an external magnetic field applied in the $z$-direction available just in two dimensions, namely $(\tau,z)$. In virtue of this dimensional reduction, QFT in a toroidal topology, investigated by Matsubara generalized formalism under an external magnetic background, was only applied in a two-dimensional momenta space in Refs.~\cite{PLAEmerson2013,MesonsMag,Tese,EPJCEmerson2017,IJMPBEmerson2018,AvanciniMesonsNeutros,PRDEmerson2019,IndianosMesonsNeutros,PhysicaAEmerson2021,PRDEmerson2022,EPJAEmerson2023,PRDEmerson2023}. 

The two dimensions apparently lost by inclusion of the magnetic effects in the propagator can be restored with some algebraic work and by taking into account integral representations for special functions, as has been done in the scalar~\cite{Lawrie,RBEFEmerson2024} and in the fermionic contexts~\cite{Chodos,Khalilov,Miransky}. The main purpose of this article is to use a hybrid method (declare here as Ritus-Schwinger method) to obtain the Green's function for both Bose and Dirac fields with no dimensional reduction and establish the Matsubara generalized formalism in the translational invariant parts of the propagation function. For this, we shall use a gauge transformation on the propagators computed. 

The paper is organized as follows: Sec II, we calculate the spinless field Green's function under a constant and homogeneous magnetic field in the $ z$-axis. This is made by a combination of Ritus' (eigenfunction method) and Schwinger (proper-time method) methods. As we will see, our expressions for Green's functions are exactly. In Sec III, we extended this hybrid method to calculate the electron propagator again under $\vec{B} = B\hat{z}$. Sec. IV is devoted to finding the expressions for the QFT in a toroidal topology without dimensional reduction imposed by the external field. We conclude the manuscript and make some remarks in Sec. V. Throughout the paper, we use $c=\hbar=k_{B}=1$ and a four-dimensional Euclidean space.    
\section{Bose Field Propagator in coordinate space under magnetic field}
\subsection{Ritus' method and the scalar propagator}

We begin by calculating the propagator of the scalar field in the Euclidean four-dimensional space under an external magnetic field $\mathbf{B}$, uniform, homogeneous, and along the $z$-direction in infinite space (bulk form). The scalar field under a magnetic background is a solution of the differential equation
\begin{eqnarray*}
\left(-D_{\mu}^{2} + m_{0}^{2}\right)\Phi\left(u\right) = 0,
\label{Eq.K-GcomCampo}
\end{eqnarray*}
where $u\equiv u^{\rho} = (\tau,x,y,z)$, $D_{\mu} ={\partial}_{\mu} - i e A_{\mu}^{ext}$, and $e$ the boson charge. Let us use the gauge: $A_{\mu}^{ext} = (0,0,xB,0)$, being $B$ a constant, such that the external magnetic field is $\vec{\nabla}\times \vec{A}^{ext} = B\hat{z}$. 

The Green's function of the scalar field satisfies the equation
\begin{eqnarray}
\left(D_{\mu}^{2} - m_{0}^{2}\right)G\left(u,u^{\prime},A\right) = -\,\delta^{4}\left(u-u^{\prime}\right).
\label{eqdeGreenBosons}
\end{eqnarray}

To find the Green function, we shall use the eigenfunction method (Ritus' method). This approach is based on the existence of a complete set of eigenfunctions $E_p(u)$ of the operator $D_{\mu}^2$. In short, the method presupposes the eigenvalue equation~\cite{Ritus,Ritus2,Mexicanos,RBEFEmerson2015,CJPEmerson2022,CMEmerson2022} 
\begin{eqnarray}
D_{\mu}^2 \,E_p = - p^2 \,E_p,
\label{eigenvalue}
\end{eqnarray}
where $E_p$ are the Ritus' eigenfunctions and
\begin{eqnarray}
D_{\mu}^2  = \nabla^{2}_{4D} - 2\,i\,\omega \,x\, \partial_{y} -\omega^{2} x^{2},
\label{operator2}
\end{eqnarray}
being $\omega \equiv e B >0$ the cyclotron frequency. Once found the complete set of eigenfunctions $E_p$, we expand the Green's function as
\begin{eqnarray}
G\left(u,u^{\prime},A\right) = \sumint dp \,E_{p}(u) \,\tilde{G}(p,A) \,E_{p}^{*}(u^{\prime}).
\label{Green}
\end{eqnarray}
Since, by construction,
\begin{eqnarray}
\sumint dp \,E_p (u) \, E^{*}_{p}(u^\prime) = \delta^4(u-u^\prime),
\label{fechamento}
\end{eqnarray}
we obtain the scalar propagator in momentum space, after using the Eqs.~(\ref{eqdeGreenBosons}), (\ref{eigenvalue}), and (\ref{Green}), namely
\begin{eqnarray}
\tilde{G}(p,A) = \frac{1}{p^2 + m_{0}^2}.
\label{propagmomenta}
\end{eqnarray}
Thus, if the eigenfunction $E_p (u)$ and the eigenvalue $p^{2}$ are found, the spinless field propagator embedded in an external magnetic field is given by Eq.~(\ref{Green}) with $\tilde{G}(p,A)$ given by Eq.~(\ref{propagmomenta}).

We solve the eigenvalue equation by the usual ansätze~\cite{Lawrie}
\begin{eqnarray}
E_p = C \exp\left[i\left(p_{\tau} \tau + p_z z + \omega \, \xi \, y\right)\right]X(x).
\label{ansatze}
\end{eqnarray}
Using the ansätze on the eigenvalue equation and the Klein-Gordon operator written in Eq.~(\ref{operator2}), we find the differential equation for $X(x)$:
\begin{eqnarray}
\left[\frac{d^{2}}{dx^{2}}-\omega^2 (x-\xi)^2 + (p^{2}-p_{\tau}^{2}-p_{z}^{2}) \right]X(x) = 0.
\label{Hermite}
\end{eqnarray}
It is well-known that the differential Hermite's equation has finite solutions only for discrete values of the independent term, such that $(p^{2}-p_{\tau}^{2}-p_{z}^{2}) \equiv \omega \, (2\,\ell+1),$ being $\ell$ non-negative integer. This fact guarantees, in Eq.(\ref{Hermite}), finite solutions.

Therefore, the solutions in variable $x$ that carry all Landau levels $\ell$ are
\begin{eqnarray}
{X}_{\ell}(x) &=& (\omega)^{1/4}\, h_{\ell}[\sqrt{\omega}(x-\xi)],           
\label{autofuncaoB}
\end{eqnarray}
where the functions $h_{\ell}[\sqrt{\omega}(x-\xi)]$ are the Hermite functions, defined as
\begin{eqnarray}
{h}_{\ell}({s}) &\equiv& \frac{1}{\sqrt{2^{\ell }\ell \,!\,\sqrt{\pi}}}\exp\left(-s^{2}/2\right) H_{\ell }\left(s\right).   
\label{FH}
\end{eqnarray}
Thanks to the Hermite polynomials $ H_{\ell }\left(s\right)$ and its properties, the functions ${h}_{\ell}(s)$ are orthonormalized~\cite{alemao}
\begin{eqnarray}
\left\lbrace
	\begin{array}{lcl}
   		\dfrac{}{}\int_{-\infty}^{\infty} \,{h}_{\ell}(s)\,{h}_{\ell^{\prime}}(s)\,ds, & = & \delta_{\ell,\ell^{\prime}}, \\
                                     \\
	\sum_{\ell=0}^{\infty} \dfrac{}{}	{h}_{\ell}(s)\,{h}_{\ell}(s^{\prime}) & = & \delta(s-s^{\prime}).
  	\end{array}
	\right.
    \label{chave}
\end{eqnarray}

By the combination of ansätze (\ref{ansatze}) for $E_{p}(u)$, second relation express in (\ref{chave}), and the complex exponential representation of Dirac delta function, we easily show the validity of  Eq.~(\ref{fechamento}), with normalization constant $|C| = 1/\sqrt{(2\pi)^{3}}$, such that the Ritus' eigenfunctions reads
\begin{eqnarray}
E_{p}(u)=\frac{\omega^{1/4}}{\sqrt{(2\pi)^3}}\,\exp\left[i\left({p_{\tau}\tau + p_{z}z} + \omega\, \xi \, y \right)\right] \, h_{\ell}[\sqrt{\omega}(x-\xi)].
\label{RitusRepaginado}
\end{eqnarray}
 
The Green's function of the zero spin field propagator under the magnetic field becomes then
\begin{eqnarray}
G\left(u,u^{\prime},A\right) =\,\sum_{\ell=0}^{\infty}\, \int \, dp_\tau \,dp_z \,\omega \, d\xi   \, E_{p}(u) \,\tilde{G}(p,A)\,E_{p}^{*}(u^{\prime}),
\label{propRitus}
\end{eqnarray}
where $E_p (u)$ is given in the Eq.~(\ref{RitusRepaginado}) and $\tilde{G}(p,A)$ is given by Eq.~(\ref{propagmomenta}) with dimensional reduction on the plane ($\mathrm{xy}$) expressed by values of $p^{2}$:
\begin{eqnarray}
p^2 = p_\tau^2+p_z^2+\omega(2\ell+1),\,\,\,\,\,\,\ell = 0,1,2,3,\cdots.
\label{pdois}
\end{eqnarray}
Physically, the quantization of orbital angular momentum is represented by the Landau levels $\ell$. 
\subsection{Getting scalar propagator expression in the proper-time representation}

Let us obtain the scalar propagator through Schwinger's proper time representation, denoted by $S$ parameter. Firstly, by using the operatorial identity
\begin{eqnarray}
\mathcal{O}^{-1} = \int_{0}^{\infty} \, dS \exp{\left(-S \, \mathcal{O}\right)},
\label{Operatorial}
\end{eqnarray}
on the Eq.~($\ref{propRitus}$), for $\mathcal{O} \equiv \left[p_{\tau}^{2}+p_{z}^{2}+\omega(2\ell+1)+m_{0}^{2}\right]$, we get
\begin{eqnarray}
G\left(u,u^{\prime},A\right) &=& 
\int_{0}^{\infty}  dS \,\int \frac{dp_\tau}{2\pi} \frac{dp_z}{2\pi} \frac{(\omega)^{3/2}}{2\pi} \,\exp\left\{i\left[p_{\tau}\left(\tau-\tau^{\prime}\right)+p_z\left(z-z^{\prime}\right)\right]\right\} \nonumber \\
&\times&\,\exp\left[-S \left(p_{\tau}^{2}+p_{z}^{2}+m_{0}^{2}\right)\right]\,\mathrm{\mathcal{I}},
\label{chaves}
\end{eqnarray}
where ${\mathcal{I}}$ is define by
\begin{eqnarray*}
\mathrm{\mathcal{I}} \equiv \,\sum_{\ell=0}^{\infty}\,\exp\left[-S\,\omega \,(2\ell+1)\right]\,\int_{-\infty}^{+\infty}\,d\xi\,\exp\left\{i\left[\omega\,\xi\left(y-y^{\prime}\right)\right]\right\} \, h_{\ell}\left[\sqrt{\omega}(x-\xi)\right]h_{\ell}\left[\sqrt{\omega}(x^{\prime}-\xi)\right].
\end{eqnarray*}
Considering the definition of Hermite functions written in Eq.~(\ref{FH}), we obtain, after completing the square in the variable $\xi$
 \begin{eqnarray}
\mathrm{\mathcal{I}} =\sum_{\ell=0}^{\infty}\,\exp\left[-S\,\omega \,(2\ell+1)\right]\,\frac{1}{2^{\ell}\,\ell!\,\sqrt{\pi}} \int_{-\infty}^{+\infty}\,d\xi\,\exp\left(\mathcal{D}\right) \, H_{\ell}\left[\sqrt{\omega}(x-\xi)\right]\,H_{\ell}\left[\sqrt{\omega}(x^{\prime}-\xi)\right], \nonumber \\
\label{Xi}
\end{eqnarray}
where $\mathcal{D}$ is short notation for
\begin{eqnarray*}
\mathcal{D} = -\omega \left\{ \left[\frac{}{}\xi-\mathcal{A}\right]^{2} +\frac{1}{4}\left[(x-x^{\prime})^{2}+(y-y^{\prime})^{2}\right] -\frac{i}{2}(x+x^{\prime})(y-y^{\prime}) \right\}\,\,;\,\,\mathcal{A}\equiv \frac{(x+x^{\prime})+i(y-y^{\prime})}{2}.
\end{eqnarray*}
Doing $\xi \,\rightarrow \, - \,\xi $, Eq.~(\ref{Xi}) reads
\begin{eqnarray}
\mathrm{\mathcal{I}} &=&\sum_{\ell=0}^{\infty}\,\exp\left[-S\,\omega \,(2\ell+1)\right]\,\frac{1}{2^{\ell}\,\ell!\,\sqrt{\pi}} \, \exp\left\{i\frac{\omega}{2}\left[\left(x+x^{\prime}\right)\left(y-y^{\prime}\right)\right]\right\}  \nonumber \\
&\times&
\exp\left\{-\frac{\omega}{4}\left[(x-x^{\prime})^{2}+(y-y^{\prime})^{2}\right]\right\} \,\int_{-\infty}^{+\infty}\,d\xi\,\exp\left[-\omega \left(\xi+\mathcal{A}\right)^{2}\right] \,\nonumber \\
&\times& H_{\ell}\left[\sqrt{\omega}(\xi+x)\right]\,H_{\ell}\left[\sqrt{\omega}(\xi+x^{\prime})\right].
\label{Xi2}
\end{eqnarray}
Making the variable change $\Theta = \sqrt{\omega}(\xi+\mathcal{A})$, we have 
\begin{eqnarray}
\mathrm{\mathcal{I}} &=&\sum_{\ell=0}^{\infty}\,\exp\left[-S\,\omega \,(2\ell+1)\right]\,\frac{1}{\sqrt{\omega} \,2^{\ell}\,\ell!\,\sqrt{\pi}} \, \exp\left\{i\frac{\omega}{2}\left[\left(x+x^{\prime}\right)\left(y-y^{\prime}\right)\right]\right\}\nonumber \\
&\times&
 \,\exp\left\{-\frac{\omega}{4}\left[(x-x^{\prime})^{2}+(y-y^{\prime})^{2}\right]\right\} \,\int_{-\infty}^{+\infty}\,d\Theta\,\exp\left(-\Theta^{2}\right) \nonumber \\
 &\times&H_{\ell}\left[\Theta+\sqrt{\omega}(x-\mathcal{A})\right]\,H_{\ell}\left[\Theta+\sqrt{\omega}(x^{\prime}-\mathcal{A})\right].
\label{Xi3}
\end{eqnarray}
Finally, using the well-known relation between the Hermite and Laguerre associated polynomials, 
\begin{eqnarray*}
\int_{-\infty}^{+\infty}\,d\Theta\,\exp\left(-\Theta^{2}\right) \, H_{\ell}\left(\Theta+a\right)\,H_{\ell^{\prime}}\left(\Theta+a^{\prime}\right) = 2^{\ell^{\prime}}\,\sqrt{\pi} \,\ell\,!(a^{\prime})^{\ell-\ell^{\prime}}\,L_{\ell}^{\ell^{\prime} - \ell}\left(-2\,a\,a^{\prime}\right),
\end{eqnarray*}
we obtain
\begin{eqnarray}
\mathrm{\mathcal{I}} &=&\,\frac{1}{\sqrt{\omega}}\,\exp\left\{i\frac{\omega}{2}\left[\left(x+x^{\prime}\right)\left(y-y^{\prime}\right)\right]\right\} \,\exp\left\{-\frac{\omega}{4}\left[(x-x^{\prime})^{2}+(y-y^{\prime})^{2}\right]\right\} \nonumber \\
&\times&\sum_{\ell=0}^{\infty}\,\exp\left[-S\,\omega \,(2\ell+1)\right]\, L^{0}_{\ell}\left[-2\,\omega\,(x-\mathcal{A})(x^{\prime}-\mathcal{A})\right].
\label{Xi4}
\end{eqnarray}
It is easy to show that
\begin{eqnarray}
(x-\mathcal{A})(x^{\prime}-\mathcal{A}) = -\frac{1}{4}\left[(x-x^{\prime})^{2}+(y-y^{\prime})^{2}\right] \equiv  -\frac{1}{4}\left[(\bf{r} - \bf{r^{\prime}})^{2}_{\perp}\right],
\end{eqnarray}
which gives
\begin{eqnarray}
\mathrm{\mathcal{I}} &=&\frac{1}{\sqrt{\omega}}\,\exp\left[i\,e\,\Phi(\bf{r}_{\perp},\bf{r}_{\perp}^{\prime})\right]\sum_{\ell=0}^{\infty}\,\exp\left[-S\,\omega \,(2\ell+1)\right]  \,\exp\left\{-\frac{\omega}{4}\left[(\bf{r} - \bf{r^{\prime}})^{2}_{\perp}\right]\right\}\, L_{\ell}\left[\frac{\omega}{2}\,(\bf{r} - \bf{r^{\prime}})^{2}_{\perp}\right], \nonumber \\
\label{Xi5}
\end{eqnarray}
where we defined the so-called Schwinger phase as
\begin{eqnarray}
\Phi({\bf{r}_{\perp}},{\bf{r}_{\perp}^{\prime}}) \equiv    \frac{B}{2}\left[\left(x+x^{\prime}\right)\left(y-y^{\prime}\right)\right]. 
\label{phase}
\end{eqnarray}

Let us rewrite Eq.~(\ref{Xi5}) in terms of the momentum coordinates as follows. There is an integral representation concerning the Gaussian multiplied by Laguerre polynomials, namely
\begin{eqnarray}
 \frac{(\alpha - \beta)^{n}}{\alpha^{n+1}}\exp\left(-{\mathcal{Y}}^{2}/2\alpha\right) L_{n}\left[\frac{\beta{\mathcal{Y}^2}}{2\alpha (\beta - \alpha)}\right]=\int_{0}^{\infty} d\mathcal{X} \mathcal{X} \exp\left(-\alpha {\mathcal{X}}^2 / 2\right) L_{n}\left(\beta{{\mathcal{X}}^{2}}/2\right) J_{0}(\mathcal{XY}), \nonumber \\
\label{Gradh}
\end{eqnarray}
where $\mathcal{Y} > 0 $, $\Re e (\alpha) > 0$ and $J_0$ is the Bessel function [\,see Ref.\cite{Grad}, equation {\bf{ET II 13(4)a}}\,]. 

We can rescale Eq.~(\ref{Gradh}) in an appropriate form: we define $\mathcal{X} \equiv p_{\perp} = \sqrt{p_x^2 + p_y^2} $, $\mathcal{Y}^{2} \equiv ({\bf{r} - \bf{r^{\prime}}})_{\perp}^2 $ and $ \alpha = \beta/2 \equiv 2/\omega$. The result is 
\begin{eqnarray*}
\exp\left\{-\frac{\omega}{4}\left[(\bf{r} - \bf{r^{\prime}})^{2}_{\perp}\right]\right\}\, L_{\ell}\left[\frac{\omega}{2}\,(\bf{r} - \bf{r^{\prime}})^{2}_{\perp}\right] &=& \frac{2\,(-1)^{\ell}}{\omega} \,\int_{0}^{\infty}\,dp_{\perp}\,p_{\perp}\, \exp\left(-\,p_{\perp}^{2}/\omega\right)\nonumber \\
&\times& L_{\ell}\left(2\,{p^{2}_{\perp}}/{\omega}\right) \,J_{0}\left(p_{\perp}\sqrt{({\bf{r} - \bf{r^{\prime}}})_{\perp}^{2}}\right).
\end{eqnarray*}
Using this result in Eq.~(\ref{Xi5}), we find
\begin{eqnarray}
\mathrm{\mathcal{I}} &=&\exp\left[i\,e\,\Phi(\bf{r}_{\perp},\bf{r}_{\perp}^{\prime})\right]\,\left(\frac{2\,\exp(-S\,\omega)}{\sqrt{\omega^{3}}}\right)\,\int_{0}^{\infty}\,dp_{\perp}\,p_{\perp}\,\exp\left(-\,p_{\perp}^{2}/\omega\right)\,J_{0}\left(p_{\perp}\sqrt{({\bf{r} - \bf{r^{\prime}}})_{\perp}^{2}}\right) \, \nonumber \\
&\times&\sum_{\ell=0}^{\infty}\,\left[-\exp\left(-\,2\,S\, \omega\right)\right]^{\ell} \, L_{\ell}\left(2\,{p^{2}_{\perp}}/{\omega}\right).
\label{Xi6}
\end{eqnarray}

Now, we are going to sum over all Landau Levels. The generating function for Laguerre polynomials is given by [\,see Ref.~\cite{Grad}, equation {\bf{EH II 189(17), MO 109}}\,]
\begin{eqnarray*}
\sum_{\ell=0}^{\infty} \mathcal{Z}^{\ell} L_{\ell}^{\epsilon}(\mathcal{Q}) = \frac{1}{(1-\mathcal{Z})^{1+\epsilon}} \exp\left[\frac{\mathcal{Q}\,\mathcal{Z}}{(\mathcal{Z}-1)}\right],
\end{eqnarray*} 
for $|\mathcal{Z}| < 1$. In the present case, we have $\mathcal{Z}\equiv - \exp[-2 \,\omega \, S]$ and $\mathcal{Q} \equiv 2\,{p^{2}_{\perp}}/{\omega}$. Therefore, the summation on $\ell$ reads (for $\epsilon = 0$),
\begin{eqnarray}
\sum_{\ell=0}^{\infty} \, \left[- \exp\left(- 2\,\omega \,S \right)\right]^{\ell} \,L_{\ell}\left(2\,{p^{2}_{\perp}}/{\omega}\right) = \frac{\exp(\omega \, S)}{2\cosh(\omega \, S)} \, \exp\left\{\frac{p^{2}_{\perp}}{\omega}\left[\frac{2}{1+\exp(2\,\omega \,S)}\right]\right\}.
\label{sumL1}
\end{eqnarray}

Finally, replacing Eq.~(\ref{sumL1}) on Eq.~(\ref{Xi6}), we get
\begin{eqnarray}
\mathrm{\mathcal{I}} &=&\frac{\exp\left[i\,e\,\Phi(\bf{r}_{\perp},\bf{r}_{\perp}^{\prime})\right]}{\sqrt{\omega^{3}} \,\cosh(\omega S)}\,\int_{0}^{\infty}\,dp_{\perp}\,p_{\perp}\,J_{0}\left(p_{\perp}\sqrt{({\bf{r} - \bf{r^{\prime}}})_{\perp}^{2}}\right) \, \nonumber \\
&\times&\exp\left[-\frac{\,p_{\perp}^{2}}{\omega}\,\tanh{(\omega S)}\right].
\label{Xi7}
\end{eqnarray}
Lastly, we are going to use the Bessel function $J_0$ in the integral representation,
\begin{eqnarray}
J_0 \left(p_{\perp}\sqrt{({\bf{r}-\bf{r^{\prime}}})^{2}_{\perp}}\right) = \frac{1}{2\pi}\int_{0}^{2\pi}d\varphi\exp\left[i p_{\perp}\sqrt{({\bf{r}-\bf{r^{\prime}}})^{2}_{\perp}} \cos\varphi\right],
\label{Bessel}
\end{eqnarray}
which one allowed us to write Eq.~(\ref{Xi7}) as
\begin{eqnarray}
\mathrm{\mathcal{I}} &=&2\pi\,\frac{\exp\left[i\,e\,\Phi(\bf{r}_{\perp},\bf{r}_{\perp}^{\prime})\right]}{\sqrt{\omega^{3}} \,\cosh(\omega S)}\,\int\,\frac{dp_{x}}{2\pi}\frac{dp_{y}}{2\pi}\, \exp\{i\left[p_{x}(x-x^{\prime})+p_{y}(y-y^{\prime})\right]\}\, \nonumber \\
&\times&\exp\left[-(p_{x}^{2}+p_{y}^{2})\,\frac{\tanh{(\omega S)}}{\omega}\right],
\label{Xi8}
\end{eqnarray}
where we have passed from cylindrical to Cartesian momenta integral. 

Since we have found $\mathrm{\mathcal{I}}$, we can replace Eq.~(\ref{Xi8}) in Eq.~(\ref{chaves}), such that the spinless propagator under a constant external magnetic field in the proper time representation is given by
\begin{eqnarray}
G\left(u,u^{\prime},A\right) &=& \exp\left[i\,e\,\Phi(\bf{r}_{\perp},\bf{r}_{\perp}^{\prime})\right]\,\int_{0}^{\infty}  dS \,\int \,\frac{d^{4}p}{(2\pi)^{4}}  \,\exp\left[ip_{\mu}\left(u-u^{\prime}\right)_{\mu}\right] \nonumber \\
&\times&\frac{1}{\cosh{(\omega S)}}\,\exp\left\{-S \left[p_{\tau}^{2}+p_{z}^{2}+(p_{x}^{2}+p_{y}^{2})\frac{\tanh{(\omega S)}}{\omega S}+m_{0}^{2}\right]\right\}.
\label{chaves55}
\end{eqnarray}
The Eq.~(\ref{chaves55}) was obtained by another method in Ref.~\cite{Lawrie} (considering the kinetic term of the Hamiltonian) and in the Minkowski space-time in Ref.~\cite{RBEFEmerson2024}. The Feynman propagator taking into account all Landau levels for the zero spin field given by Eq.~(\ref{chaves55}), gives exactly the free propagator expression when we take the limit $\omega \rightarrow 0$ in Eq.~(\ref{chaves55}).

Notice that the scalar propagator under the magnetic field along the $z$-direction is not invariant under translation on the orthogonal plane to the external field. This is due to the Schwinger phase $\Phi(\bf{r}_{\perp},\bf{r}_{\perp}^{\prime})$. However, we can make a gauge transformation to eliminate the translational non-invariant part of Green's function as follows. 

Initially, we will show that the translationally noninvariant part of the propagator can be obtained by integrating the vector potential along a straight line connecting the initial and final points in the orthogonal plane to the magnetic field, i.e.,
\begin{eqnarray}
\Phi(\mathbf{r}_{\perp},\mathbf{r}_{\perp}^{\prime}) = \int_{\mathbf{r}_{\perp}^{\prime}}^{\mathbf{r}_{\perp}}\,\vec{A}(\vec{\xi}\,)\cdot d{\vec{\xi}},
\end{eqnarray}
where
\begin{eqnarray*}
\vec{A}(\vec{\xi}\,)  = \left(0,B\xi_{1},0\right) \,\,\,\Longrightarrow\,\,\vec{\nabla}_{\vec{\xi}} \times \vec{A}(\vec{\xi}\,)={B}\hat{z}.
\end{eqnarray*}
We can parameterize the straight line in terms of a parameter $\lambda$ (see chapter $3$ of the Ref.~\cite{LivroSch})
\begin{eqnarray*}
\vec{\xi}  = (1-\lambda)\,\vec{r}_{\perp}^{\,\prime}+\lambda\,\vec{r}_{\perp},\,\,\,\,\,\therefore\,\,\,\,\, d\vec{\xi} = (\vec{r}_{\perp}-\vec{r}_{\perp}^{\,\prime})\,d\lambda, \,\,\,\,\,\,0\leq\lambda\leq1.
\end{eqnarray*}
Thus,
\begin{eqnarray}
\int_{\mathbf{r}_{\perp}^{\prime}}^{\mathbf{r}_{\perp}}\,\vec{A}(\vec{\xi}\,)\cdot d{\vec{\xi}} &=& \int_{0}^{1}\,B\left[(1-\lambda)\,x^{\prime}+\lambda \,x\right](y-y^{\prime})\,d\lambda \nonumber\\
&=&\frac{B}{2}(x+x^{\prime{}})(y-y^{\prime})\equiv\Phi(\mathbf{r}_{\perp},\mathbf{r}_{\perp}^{\prime}).
\end{eqnarray}

We see that, can write Eq.~(\ref{chaves55}) as
\begin{eqnarray}
G\left(u,u^{\prime},A\right) &=& \exp\left[i\,e\int_{\mathbf{r}_{\perp}^{\prime}}^{\mathbf{r}_{\perp}}\,\vec{A}(\vec{\xi}\,)\cdot d{\vec{\xi}}\,\right]\,\int_{0}^{\infty}  dS \,\int \,\frac{d^{4}p}{(2\pi)^{4}}  \,\exp\left[ip_{\mu}\left(u-u^{\prime}\right)_{\mu}\right] \nonumber \\
&\times&\frac{1}{\cosh{(\omega S)}}\,\exp\left\{-S \left[p_{\tau}^{2}+p_{z}^{2}+(p_{x}^{2}+p_{y}^{2})\frac{\tanh{(\omega S)}}{\omega S}+m_{0}^{2}\right]\right\}.
\label{chaves56}
\end{eqnarray}

Since we have the freedom to choose an arbitrary function $\Lambda(\vec{\xi})$ such that it does not generate any influence on the external field, we obtain an equivalent expression for the propagation function according to the prescription
\begin{eqnarray}
 \vec{A} \, \mapsto \, \vec{A} + \vec{\nabla} \, \Lambda \, \Longrightarrow \,  G\left(u,u^{\prime},A\right)  \, \mapsto \, G\left(u,u^{\prime},A+\partial \Lambda\right) . 
 \label{gauge4}
\end{eqnarray}

Without loss of generality, we can choose the arbitrary function as $\Lambda(\vec{\xi}\,) = -\,{B(x+x^{\prime})\xi_{2}}/{2}$, which has gradient integrated over the straight line given by
\begin{eqnarray}
\int_{\mathbf{r}_{\perp}^{\prime}}^{\mathbf{r}_{\perp}}\,\vec{\nabla}\,\Lambda(\vec{\xi}\,)\cdot d{\vec{\xi}} = -\frac{B}{2}(x+x^{\prime})(y-y^{\prime}).
\end{eqnarray}

Therefore, through gauge transformation written in Eq.~(\ref{gauge4}), the Green's function $G\left(u,u^{\prime},A+\partial \Lambda\right)$ of the spinless field under a magnetic background becomes independent of gauge, namely
\begin{eqnarray}
G\left(u-u^{\prime},\omega\right) &=& 
\int_{0}^{\infty}  dS \,\int \,\frac{d^{4}p}{(2\pi)^{4}}  \,\exp\left[ip_{\mu}\left(u-u^{\prime}\right)_{\mu}\right]\nonumber \\
&\times&\frac{1}{\cosh{(\omega S)}}\,\exp\left\{-S \left[p_{\tau}^{2}+p_{z}^{2}+(p_{x}^{2}+p_{y}^{2})\frac{\tanh{(\omega S)}}{\omega S}+m_{0}^{2}\right]\right\}.
\label{chaves57}
\end{eqnarray}
Additionally, to the fact that Eq.~(\ref{chaves57}) is independent of the chosen gauge, it is invariant under translation and will be used to establish the Matsubara generalized formalism in section IV.
\section{Dirac Field Propagator in coordinate space under magnetic field}
\subsection{Ritus' method and the fermion propagator}

The Dirac equation in the four-dimensional Euclidean space under a magnetic field along $z$ direction is provided by
\begin{eqnarray*}
({\slashed{D}} +i\,m_{0} )\Psi(u) = 0,
\label{Eq. Dirac}
\end{eqnarray*}
where again we have use the minimal coupling $D_{\mu} = \partial_{\mu}-ie{A}_{\mu}^{ext}$ and $A_{\mu}^{ext} = (0,0,xB,0)$. The Green's function in this fermionic case is given by 
\begin{eqnarray}
({\slashed{D}} +i \,m_{0} ) \,\mathbb{S}(u,u^{\prime},A) =  \mathbb{I}_{4}\,\delta^{4}(u-u^{\prime}).
\label{eq. de Green}
\end{eqnarray}
We are using the gamma matrices in Euclidean space defined as~\cite{CJPEmerson2022}
\begin{equation}
\mathbb{\gamma}_{\mathrm{0}} =-i\left( 
\begin{array}{cc}
\mathrm{0}_{2} & \mathbb{I}_{2} \\ 
\mathbb{I}_{2} & \mathrm{0}_{2}%
\end{array}\right) \,;\,
\mathbb{\gamma}_{\mathrm{j}} =\left( 
\begin{array}{cc}
0_{2} & -\mathbb{\sigma}_{{j}} \\ 
\mathbb{\sigma}_{{j}} & 0_{2}%
\end{array}\right), \,\,\,\,{j=1,2,3.}
\label{matrizes de Dirac}
\end{equation}
being $\mathbb{\sigma}_{j}$ the Pauli matrices. According to this representation, we have
\begin{eqnarray*}
{\slashed{D}} &=&\gamma_{\mu}D_{\mu} \ ; \ \{\gamma_{\mu},\gamma_{\nu}\} =-\,2\mathrm{\delta}_{\mu\nu}.
\end{eqnarray*}

The commutation of $[{\slashed{D}}^{2},\slashed{D}] = 0$ allows us to look for eigenfunctions of the quadratic Dirac operator. Then, let us write Eq.~(\ref{eq. de Green}) as
\begin{eqnarray}
({\slashed{D}}^{2} + \,m_{0}^{2} \, ) \,\mathbb{G}(u,u^{\prime},A) = \mathbb{I}_{4}\, \delta^{4}(u-u^{\prime}),
\label{aa}
\end{eqnarray}
being $\mathbb{G}(u,u^{\prime},A)$ an auxiliary propagator, related by the electron propagator $\mathbb{S}(u,u^{\prime},A)$, by
\begin{eqnarray}
\mathbb{S} (u,u^{\prime},A) \equiv ({\slashed{D}} -i\, \,m_{0} \, ) \,\mathbb{G}(u,u^{\prime},A).
\label{eq. de Green MM}
\end{eqnarray}
Therefore, let us compute initially the quantity $\mathbb{G}(u,u^{\prime},A)$ and after this, applying the operator $({\slashed{D}} -i\, \,m_{0} \, )$ over $\mathbb{G}$ to obtain the fermionic propagator $\mathbb{S} (u,u^{\prime},A)$.

By the definitions above, it is easy to show that
\begin{eqnarray*}
	{\slashed{D}}^{2} = \frac{1}{4}\left(\frac{}{}i e [\gamma_{\mu},\gamma_{\nu}]F_{\mu\nu}-4 D_{\mu}^{2}\,\right),
\end{eqnarray*}
where $D_{\mu}^{2}$, is the same operator written in Eq.~(\ref{operator2}). For a pure magnetic background, we have $F_{12} = - F_{21} = B$. The Dirac matrices defined in Eq.~(\ref{matrizes de Dirac}) allow us write 
\begin{eqnarray}
	\slashed{D}^{2} = \omega \left(\mathbb{I}_{2}\otimes \mathbb{\sigma}_{z}\right)- D_{\mu}^{2},
    \label{Aila}
\end{eqnarray}
where
\begin{equation}
\left(\mathbb{I}_{2}\otimes\sigma_{z}\right) =\left( 
\begin{array}{cccc}
1 & 0 & 0 & 0\\ 
0 & -1 & 0 & 0\\
0 & 0 & 1 & 0\\
0 & 0 & 0 & -1
\end{array}\right).
\label{matriz}
\end{equation}

As we have done in the bosonic case, if we find the eigenfunctions $\mathbb{E}_{p}$ of the quadratic Dirac operator, we can write the auxiliary propagator $\mathbb{G}$ by expansion in terms of them, namely
\begin{eqnarray}
\mathbb{G}\left(u,u^{\prime},A\right) =\sumint \,dp \, {\mathbb{E}}_{p}(u) \,\tilde{\mathbb{G}}(p,A)\,{\mathbb{E}}_{p}^{\dagger}(u^{\prime}).
\label{43} 
\end{eqnarray}
From Eq.~(\ref{Aila}) we observe that the only difference between the scalar propagator $G$ and the auxiliary fermion propagator $\mathbb{G}$ is the spin term $(\mathbb{I}_{2}\otimes\sigma_{z})$. Thus, it is sufficient we define the matrix ansätze
\begin{eqnarray}
{\mathbb{E}}_{p}(u) = \,E_{p}(u)\,{\mathbf{\Omega}}_{\mathrm{\sigma}_{\mathrm{B}}}\,,
\label{ansatzenova}
\end{eqnarray}
being the matrix ${\mathbf{\Omega}}_{\mathrm{\sigma}_{\mathrm{B}}}$ diagonal
\begin{eqnarray}
{\bf{\Omega}}_{\sigma_{\rm{B}}} &=& {\bf{diag}}\left(\frac{}{}\delta_{1,\sigma_{\rm{B}}}\,;\,\delta_{-1,\sigma_{\rm{B}}}\,; \,\delta_{1,\sigma_{\rm{B}}}\,;\,\delta_{-1,\sigma_{\rm{B}}}\frac{}{}\right),
\label{diag}
\end{eqnarray}
and $E_{p}(u)$ are given by Eq.~(\ref{RitusRepaginado}).

Note that the matrix ${\mathbb{E}}_{p}(u)$ satisfies
\begin{eqnarray}
\left(\mathbb{I}_{2}\otimes\mathbb{\sigma}_{z}\right){\mathbb{E}}_{p}(u)&=&{\mathbb{E}}_{p}(u)\left(\mathbb{I}_{2}\otimes\mathbb{\sigma}_{z}\right).
\label{ansatze5}
\end{eqnarray}

From Eqs.~(\ref{ansatzenova}), (\ref{diag}) and  (\ref{fechamento}), besides the relation $\left({\mathbf{\Omega}}_{\mathrm{\sigma}_{\mathrm{B}}} \,{\mathbf{\Omega}}^{\mathrm{T}}_{\mathrm{\sigma}_{\mathrm{B}}} \right) = \mathbb{I}_{4}$, we easily show that
\begin{eqnarray}
\sumint \, dp \, {\mathbb{E}}_{p}(u) \,{\mathbb{E}}_{p}^{\dagger}(u^{\prime}) = \mathbb{I}_{4}\,\delta^{4}(u-u^{\prime}).
\label{propRitusF} 
\end{eqnarray}

Performing analogous steps to the bosonic case, i.e., replacing Eq.~(\ref{Aila}), (\ref{43}), (\ref{ansatzenova}) on (\ref{aa}), and considering the representation of Dirac delta written in Eq.~(\ref{propRitusF}) together with the property noted in equation in (\ref{ansatze5}), we found the electron propagator in momenta space, given by the matrix structure 
\begin{eqnarray}
\tilde{\mathbb{G}}(p,A)=\frac{1}{\left[p^{2}+m_{0}^{2}+\omega\left(\mathbb{I}_{2}\otimes\mathbb{\sigma}_{z}\right)\right]},
\label{propRitusFF} 
\end{eqnarray}
where $p^2$ is defined on Eq.~(\ref{pdois}).

Now, we can use Eq.~(\ref{43}) to obtain the fermion propagator under a magnetic field along $z$ direction. 
\subsection{Getting fermionic propagator expression in the proper-time representation}

In this part, we use the Schwinger trick given by Eq.~(\ref{Operatorial}) and perform the summation over the Landau levels $\ell$. The computations are the same as the lead us Eq.~(\ref{chaves55}), except for the spin term. The result is
\begin{eqnarray}
\mathbb{G}\left(u,u^{\prime},A\right) &=& \exp\left[i\,e\,\Phi(\bf{r}_{\perp},\bf{r}_{\perp}^{\prime})\right]\,\int_{0}^{\infty}  dS \,\int \,\frac{d^{4}p}{(2\pi)^{4}}  \,\exp\left[ip_{\mu}\left(u-u^{\prime}\right)_{\mu}\right]\,\frac{1}{\cosh{(\omega S)}}\nonumber \\
&\times&\exp\left\{-S \left[p_{\tau}^{2}+p_{z}^{2}+(p_{x}^{2}+p_{y}^{2})\frac{\tanh{(\omega S)}}{\omega S}+m_{0}^{2}+\omega\left(\mathbb{I}_{2}\otimes\mathbb{\sigma}_{z}\right)\right]\right\},
\label{chaves65}
\end{eqnarray}
where the Schwinger's phase was written in Eq.~(\ref{phase}). 

Finally, let us applying the operator $({\slashed{D}} -i\, \,m_{0} \, )$ over Eq.~(\ref{chaves65}) to obtain the fermionic propagator $\mathbb{S} (u,u^{\prime},A)$ according to Eq.~(\ref{eq. de Green MM}). The application is immediately and results in
\begin{eqnarray}
\mathbb{S}\left(u,u^{\prime},A\right) &=& \exp\left[i\,e\,\Phi(\bf{r}_{\perp},\bf{r}_{\perp}^{\prime})\right]\,\int_{0}^{\infty}  dS \,\int \,\frac{d^{4}p}{(2\pi)^{4}} \left\{i\slashed{p} - im_{0} +i\frac{\omega}{2}\left[\gamma_{x}(y-y^{\prime})-\gamma_{y}(x-x^{\prime})\right]\right\} \nonumber \\
&\times&\frac{\exp\left[ip_{\mu}\left(u-u^{\prime}\right)_{\mu}\right]}{\cosh{(\omega S)}} \,\exp\left\{-S \left[p_{\tau}^{2}+p_{z}^{2}+(p_{x}^{2}+p_{y}^{2})\frac{\tanh{(\omega S)}}{\omega S}+m_{0}^{2}+\omega\left(\mathbb{I}_{2}\otimes\mathbb{\sigma}_{z}\right)\right]\right\}. \nonumber \\
\label{chaves66}
\end{eqnarray}
However, we can write typical terms on Eq.~(\ref{chaves66}) as
\begin{eqnarray}
& &\frac{d}{dp_{x}}\left\{\frac{}{}\exp\left[ip_{x}(x-x^{\prime})\right] \, \exp\left[-p_{x}^{2} \tanh{(\omega \,S)} / \omega \right]\frac{}{}\right\} = \nonumber \\
& &\left[ i(x-x^{\prime}) - \frac{2p_{x}}{\omega}\tanh{(\omega S)} \right] \left\{\frac{}{}\exp\left[ip_{x}(x-x^{\prime})\right] \, \exp\left[-p_{x}^{2} \tanh{(\omega \,S)} / \omega \right]\frac{}{}\right\}.
\label{int}
\end{eqnarray}
Integrating term by term in Eq.~(\ref{int}) in variable $-\infty \leq p_{x} \leq +\infty$, and considering the decreasing factor in $\exp{(-p_{x}^{2})}$ at the border of the domain, we obtain
\begin{eqnarray}
&&\int_{-\infty}^{+\infty}\,\frac{dp_{x}}{2\pi}\,\left[i(x-x^{\prime})\right]\left\{\frac{}{}\exp\left[ip_{x}(x-x^{\prime})\right] \, \exp\left[-p_{x}^{2} \tanh{(\omega \,S)} / \omega \right]\frac{}{}\right\} = \nonumber \\
&&\int_{-\infty}^{+\infty}\,\frac{dp_{x}}{2\pi}\,\frac{2p_{x}}{\omega}\,\tanh{(\omega S)}\left\{\frac{}{}\exp\left[ip_{x}(x-x^{\prime})\right] \, \exp\left[-p_{x}^{2} \tanh{(\omega \,S)} / \omega \right]\frac{}{}\right\}.
\label{1}
\end{eqnarray}
Similarly, for the $y$ coordinate
\begin{eqnarray}
&&\int_{-\infty}^{+\infty}\,\frac{dp_{y}}{2\pi}\,\left[i(y-y^{\prime})\right]\left\{\frac{}{}\exp\left[ip_{y}(y-y^{\prime})\right] \, \exp\left[-p_{y}^{2} \tanh{(\omega \,S)} / \omega \right]\frac{}{}\right\} = \nonumber \\
&&\int_{-\infty}^{+\infty}\,\frac{dp_{y}}{2\pi}\,\frac{2p_{y}}{\omega}\,\tanh{(\omega S)}\left\{\frac{}{}\exp\left[ip_{y}(y-y^{\prime})\right] \, \exp\left[-p_{y}^{2} \tanh{(\omega \,S)} / \omega \right]\frac{}{}\right\}
\label{2}
\end{eqnarray}
Replacing Eqs.~(\ref{1}) and (\ref{2}) on Eq.~(\ref{chaves66}), we get
\begin{eqnarray}
\mathbb{S}\left(u,u^{\prime},A\right) &=& \exp\left[i\,e\,\Phi(\bf{r}_{\perp},\bf{r}_{\perp}^{\prime})\right]\,\int_{0}^{\infty}  dS \,\int \,\frac{d^{4}p}{(2\pi)^{4}} \left\{i\slashed{p} - im_{0} +\left[\gamma_{x}p_{y}-\gamma_{y}p_{x}\right]\,\tanh{(\omega S)}\right\} \nonumber \\
&\times&\frac{\exp\left[ip_{\mu}\left(u-u^{\prime}\right)_{\mu}\right]}{\cosh{(\omega S)}} \,\exp\left\{-S \left[p_{\tau}^{2}+p_{z}^{2}+(p_{x}^{2}+p_{y}^{2})\frac{\tanh{(\omega S)}}{\omega S}+m_{0}^{2}+\omega\left(\mathbb{I}_{2}\otimes\mathbb{\sigma}_{z}\right)\right]\right\}. \nonumber \\
\label{chaves67}
\end{eqnarray}

Using the series expansion on the spin term,
\begin{eqnarray*}
\exp{\left[-\omega\,S\left(\mathbb{I}_{2}\otimes\sigma_{z}\right)\right]} = \sum_{n=0}^{\infty}\,\frac{\left[-\omega\,S\left(\mathbb{I}_{2}\otimes\sigma_{z}\right)\right]^{n}}{n!},
\end{eqnarray*}
we obtain
\begin{eqnarray*}
\exp{\left[-\omega\,S\left(\mathbb{I}_{2}\otimes\sigma_{z}\right)\right]} &=& \mathbb{I}_{4}\left[1+\frac{(\omega S)^{2}+}{2!}+\frac{(\omega S)^{4}+}{4!}+\cdots\right] \nonumber \\
&&-\left(\mathbb{I}_{2}\otimes\sigma_{z}\right)\left[(\omega S)+\frac{(\omega S)^{3}+}{3!}+\frac{(\omega S)^{5}}{5!}+\cdots\right],
\end{eqnarray*}
which gives
\begin{eqnarray}
\exp{\left[-\omega\,S\left(\mathbb{I}_{2}\otimes\sigma_{z}\right)\right]} = \mathbb{I}_{4}\cosh{(\omega S)}-\left(\mathbb{I}_{2}\otimes\sigma_{z}\right)\sinh{(\omega S)}.
\label{m}
\end{eqnarray}
Therefore, after replacing Eq.~(\ref{m}) on Eq.~(\ref{chaves67}), we have the electron propagator in the proper-time representation:
\begin{eqnarray}
\mathbb{S}\left(u,u^{\prime},A\right) &=& \exp\left[i\,e\,\Phi(\bf{r}_{\perp},\bf{r}_{\perp}^{\prime})\right]\,\int_{0}^{\infty}  dS \,\int \,\frac{d^{4}p}{(2\pi)^{4}} \left\{\frac{}{}i\slashed{p} - im_{0} +\left[\gamma_{x}p_{y}-\gamma_{y}p_{x}\right]\,\tanh{(\omega S)}\right\} \nonumber \\
&\times&\exp\left[ip_{\mu}\left(u-u^{\prime}\right)_{\mu}\right] \,\exp\left\{-S \left[p_{\tau}^{2}+p_{z}^{2}+(p_{x}^{2}+p_{y}^{2})\frac{\tanh{(\omega S)}}{\omega S}+m_{0}^{2}\right]\right\} \nonumber \\
&\times&\left[\mathbb{I}_{4} - \left(\mathbb{I}_{2}\otimes\sigma_{z}\right) \tanh{(\omega S)}\right].
\label{chaves68}
\end{eqnarray}
The Eq.~(\ref{chaves68}) also was obtained by Miransky and Shovkovy in equation number ($107$) of Ref.~\cite{Miransky}.

Using again the gauge transformation written in Eq.~(\ref{gauge4}), the Green's function $\mathbb{S}\left(u,u^{\prime},A+\partial \Lambda\right)$ of the fermionic field under a magnetic background becomes independent of gauge
\begin{eqnarray}
\mathbb{S}\left(u-u^{\prime},\omega\right) &=& \int_{0}^{\infty}  dS \,\int \,\frac{d^{4}p}{(2\pi)^{4}} \left\{\frac{}{}i\slashed{p} - im_{0} +\left[\gamma_{x}p_{y}-\gamma_{y}p_{x}\right]\,\tanh{(\omega S)}\right\} \nonumber \\
&\times&\exp\left[ip_{\mu}\left(u-u^{\prime}\right)_{\mu}\right] \,\exp\left\{-S \left[p_{\tau}^{2}+p_{z}^{2}+(p_{x}^{2}+p_{y}^{2})\frac{\tanh{(\omega S)}}{\omega S}+m_{0}^{2}\right]\right\} \nonumber \\
&\times&\left[\mathbb{I}_{4} - \left(\mathbb{I}_{2}\otimes\sigma_{z}\right)\tanh{(\omega S)}\right].
\label{chaves69}
\end{eqnarray}
We take notice that Eq.~(\ref{chaves69}) in the limit $B \,\rightarrow \,0$, gives the electron free propagator, as should be.
\section{QFT in a toroidal topology under an external magnetic field
}

In this section, let us obtained QFT in a toroidal topology without the dimensional reduction imposed by the external field. Initially, we fix $u^{\prime} = (0,0,0,0)$ in both scalar and spinorial Green's functions calculated in the Eqs.~(\ref{chaves57}) and (\ref{chaves69}). Such choose does not diminishes the generality of $G(u-u^{\prime},\omega)$ and $\mathbb{S}(u-u^{\prime},\omega)$, although it provides simplifications in the equations computed below.

As we have mentioned in the introduction, the KMS conditions establish, for bosons, periodic boundary conditions in the imaginary-time coordinate $\tau$
\begin{eqnarray*}
G(\tau,x_{j},\omega\,) &=& + \, \,G(\tau+\beta,x_{j},\omega\,),\,\,\,\,j=1,2,3.
\end{eqnarray*}
and, for fermions, antiperiodic boundary conditions 
\begin{eqnarray*}
\mathbb{S}(\tau,x_{j},\omega\,) &=& - \,\, \mathbb{S}(\tau+\beta,x_{j},\omega\,),\,\,\,\,j=1,2,3.
\end{eqnarray*}
where $\beta^{-1} = T$ is the temperature of heated system.

In terms of momentum coordinates, the corresponding KMS conditions read, for bosons,
\begin{eqnarray}
p_{\tau} \rightarrow  {\omega}_{n_\tau} \equiv \frac{\pi}{\beta}
\left(2  n_{\tau}\right)&-&i\mu, \,\,\,\,\,\,\, n_{\tau} = 0,\pm 1 , \pm 2, \cdots. \nonumber\\
 \int \,\frac{dp_{\tau}}{2\pi}\,f\left(p_{\tau},{p}_{j}\,\right) & \rightarrow & \frac{1}{\beta }\,\sum_{ n_{\tau}=-\infty}^{\infty} \,f \left( {\omega}_{n_{\tau}},p_{j}\,\right),\,\,\,\,j=1,2,3.
 \label{Mat1}
\end{eqnarray}
and, for fermions,
\begin{eqnarray}
p_{\tau} \rightarrow  {\omega}_{n_\tau} \equiv \frac{\pi}{\beta}
\left(2  n_{\tau}+ 1\right)&-&i\mu, \,\,\,\,\,\,\, n_{\tau} = 0,\pm 1 , \pm 2, \cdots. \nonumber\\
 \int \,\frac{dp_{\tau}}{2\pi}\,f\left(p_{\tau},p_{j}\,\right) & \rightarrow & \frac{1}{\beta }\,\sum_{ n_{\tau}=-\infty}^{\infty} \,f \left( {\omega}_{n_{\tau}},{p}_{j}\,\right),\,\,\,\,j=1,2,3.
 \label{Mat2}
\end{eqnarray}
where $\mu$ is the chemical potential of the system. 

For the inclusion of a nontrivial topology in the system, we do not need to use only periodic or antiperiodic boundary conditions in the spatial coordinates according to the nature bosonic or fermionic of the system, respectively. Actually, we can interpolate these two kinds of boundary conditions in a unified way, defined for bosons as
\begin{eqnarray*}
G \, (\tau,x_{j},\omega\,) &=& \exp{(-i\,\pi\,\alpha_{j})} \, G \, (\tau,x_{j}+L_{j},\omega\,),\,\,\,\,j=1,2,3,
\end{eqnarray*}
and for fermions as
\begin{eqnarray*}
\mathbb{S} \, (\tau,x_{j},\omega\,) &=& \exp{(-i\,\pi\,\alpha_{j})} \, \mathbb{S} \,(\tau,x_{j}+L_{j},\omega\,),\,\,\,\,j=1,2,3,
\end{eqnarray*}
where $\alpha_{j} \,=\, 0$ represents periodic boundary conditions and  $\alpha \,=\, 1$ antiperiodic boundary conditions. In the range \,$0 \, < \alpha < \, 1$\, the system will be subject to the twisted boundary conditions.

Therefore, we have the following prescription over the spatial momenta 
\begin{eqnarray}
p_{j} \rightarrow {\omega}_{n_j} \,\,\equiv\,\, \frac{\pi}{L_j}
\left(2 \, n_{j}\right)&-&i\,\mu_{j} \,\,\,\,;\,\,\,\,  \mu_{j} = i \frac{\pi}{L_j}\alpha_{j},\,\,\,\,j=1,2,3,\nonumber      \\
 \int\,\frac{dp_{j}}{2\pi}\,f(p_{\tau},p_{j}\,) & \rightarrow & \frac{1}{ L_{j}}\,\sum_{ n_{j}=-\infty}^{\infty} f \left(p_{\tau},{\omega}_{n_{j}}\,\right),\,\,\,\,j=1,2,3.
 \label{Mat22}
\end{eqnarray}
where $n_{j} = 0,\pm 1 , \pm 2, \cdots$ and $L_j$ is the length of the compactified spatial coordinate $x_{j}$. 

The Eqs.(\ref{chaves57}) and (\ref{chaves69}) together with Eqs.~(\ref{Mat1}), (\ref{Mat2}) and (\ref{Mat22}) define the Matsubara generalized prescription under the magnetic background without any dimensional reduction over spatial coordinates.
\section{Comments and Conclusions}

We have applied the Ritus-Schwinger method for calculating the Green's functions of Bose and Dirac fields under a constant and homogeneous external magnetic field along to $z$-axis. Through the Ritus' eigenfunctions and Schwinger's trick, we perform the summation over all Landau levels to get a closed expression for the scalar and spinor propagators. Furthermore, we made a gauge transformation and eliminated the part noninvariant under translation of the propagators. This made it possible to establish the Matsubara generalized prescription on the bosonic and fermionic systems embedded in an external magnetic field. Our computations does not use the additional Ritus' condition for the fermionic field, used by him in the papers of the 1970s decade. Also, we do not need to use all the machinery developed by Schwinger in his seminal paper of 1951. For its simplicity, the hybrid method discussed here is an interesting alternative to obtain the Green's function in the external backgrounds under a nontrivial topology.

%
%

\end{document}